\documentstyle[aps,prl,preprint,floats,epsfig]{revtex}

\begin{document}

\preprint{\tighten\vbox{\hbox{\hfil CLNS 97/1501}
                        \hbox{\hfil CLEO 97-17}}}

\title{\Large\bf Investigation of Semileptonic \boldmath $B$ Meson Decay to 
P-Wave Charm Mesons}  

\author{CLEO Collaboration}
\date{\today}

\maketitle
\tighten
\medskip

\baselineskip 19pt

\begin{abstract}
We have studied semileptonic $B$ meson decays with a P-wave charm
meson in the final state using $3.29 \times 10^6$\ $B\bar{B}$ events
collected by the CLEO~II detector at the Cornell Electron-positron
Storage Ring.  We find a value for the exclusive semileptonic product
branching fraction: ${{\cal B}(B^- \to D_1^0 \ell^- \bar{\nu}_{\ell})
{\cal B}(D_1^0 \to D^{*+}\pi^-)} = (0.373 \pm 0.085 \pm 0.052 \pm 0.024)\%$
and an upper limit for ${{\cal B}(B^- \to D_2^{*0} \ell^- \bar{\nu}_{\ell})
{\cal B}(D_2^{*0} \to D^{*+}\pi^-)} < 0.16\%$ (90\%~{\rm C.L.}). These 
results indicate that at least 20\% of the total
$B^-$ semileptonic rate is unaccounted for by the observed exclusive decays,
$B^- \to D^0 \ell^- \bar{\nu}$,
$B^- \to D^{*0} \ell^- \bar{\nu}$,
$B^- \to D_1^0 \ell^- \bar{\nu}$, and
$B^- \to D_2^{*0} \ell^- \bar{\nu}$.
\end{abstract}

\vspace*{3.5cm}

\centerline{(Submitted to Physical Review Letters)}

\newpage

\renewcommand{\thefootnote}{\alph{footnote}}


\begin{center}
A.~Anastassov,$^{1}$ J.~E.~Duboscq,$^{1}$ D.~Fujino,$^{1,}$%
\footnote{Permanent address: Lawrence Livermore National Laboratory, Livermore, CA 94551.}
K.~K.~Gan,$^{1}$ T.~Hart,$^{1}$ K.~Honscheid,$^{1}$
H.~Kagan,$^{1}$ R.~Kass,$^{1}$ J.~Lee,$^{1}$ M.~B.~Spencer,$^{1}$
% ...
\end{center}
 
\small
\begin{center}
$^{1}${Ohio State University, Columbus, Ohio 43210}\\
% ...
\end{center}
 
\setcounter{footnote}{0}

\newpage


\medskip

There is general agreement among a number of
measurements of the exclusive semileptonic
$\bar{B}$ meson decays,
$\bar{B} \to D \ell \bar{\nu}_{\ell}$ and
$\bar{B} \to D^* \ell \bar{\nu}_{\ell}$~\cite{breview}.
Together they account for approximately 60 -- 70\% of the inclusive
$\bar{B} \to X \ell \bar{\nu}_{\ell}$ branching fraction~\cite{jdr+prb}.  
Since the branching fraction for $b \to u \ell
\bar{\nu}$ is known to be small, the missing exclusive decays must be 
sought among $b \to c \ell \bar{\nu}$ decays to higher mass $D_J$ 
states or nonresonant hadronic states with a $D$ or $D^*$ and other 
hadrons.  Measurements of $B^-\to D_1^0\ell^-\bar{\nu}_{\ell}$ 
and $B^- \to D_2^{*0} \ell^- \bar{\nu}_{\ell}$ have
been reported previously~\cite{aleph:dss,opal:dss}.
In this paper we report new measurements of these two decay modes.

The $D_J$ mesons contain one charm quark and one light quark with relative
angular momentum $L=1$.  The quark spins can sum to $S=0$ or $S=1$, so 
there are four spin-parity states given by $J^P = 1^+$ or $0^+$, $1^+$, 
and $2^+$.  Parity and angular momentum conservation restrict the 
decays available to the four states.  According to Heavy Quark Effective 
Theory (HQET), there exists an approximate
spin-flavor symmetry for hadrons consisting of one heavy and one light
quark~\cite{hqet}.   In the limit of infinite heavy quark mass, such 
mesons are described by the total angular momentum of the light 
constituents $j=S_q+L$.  In HQET, 
the $D_J$ mesons make up two
doublets, $j = 1/2$ and $j = 3/2$.   The members of the $j = 3/2$ doublet are
predicted to decay only in a D-wave and to be relatively narrow.  The  $j = 1/2$
mesons are predicted to decay only in an S-wave and to be relatively broad.
In this analysis we study the semileptonic decays of the $B$ meson to final
states containing the narrow ($j$=3/2) excited charm  mesons: the
$^jL_J =\: ^{3/2}\!P_2$ and $^{3/2}\!P_1$, called $D_2^*$ 
and $D_1$, respectively~\cite{pdbook}.

The data used in this analysis were collected by the CLEO~II 
detector at the Cornell Electron-positron Storage Ring (CESR).  
The CLEO~II detector~\cite{detector} is a 
multipurpose high energy physics detector incorporating excellent
charged and neutral particle detection and measurement.
The data sample consists of an integrated luminosity of 3.11 fb$^{-1}$
on the $\Upsilon(4S)$ resonance (ON Resonance), corresponding 
to $3.29\times 10^6$ 
$B\bar{B}$ events, and 1.61 fb$^{-1}$ at a center-of-mass energy
$\sim$ 55 MeV below the $\Upsilon(4S)$ resonance (OFF Resonance).

The exclusive ${B^- \to D_J^0 \ell^- \bar{\nu}_{\ell}}$ decay 
is studied by reconstructing 
the decay channel $D_J^0 \to D^{*+}\pi^-$ using
the decay chain $D^{*+} \to  D^0\pi^+$, and $D^0 \to K^-\pi^+$ or $D^0
\to K^-\pi^+\pi^0$~\cite{chgconj}. Hadronic events are required to
have at least one track identified as a lepton 
with momentum between 0.8 GeV/$c$ and 2.0 GeV/$c$ for electrons
and between 1.0 GeV/$c$ and 2.0 GeV/$c$ for muons. Electrons are 
identified by matching energy deposited in the
CsI calorimeter and momentum measured in the drift chamber, their 
energy loss in the drift chamber gas and their time of flight 
in the detector.  The muon identification relies upon penetration 
through layers of iron absorber to muon chambers.  To reduce 
non-$B\bar{B}$ background (contamination of our sample by $e^+e^-$ 
interactions which result in $q\bar{q}$ hadronization rather that 
producing an $\Upsilon(4S)$ meson), each event is required 
to satisfy the ratio of Fox-Wolfram~\cite{R2} 
moments $R_2<0.4$.  All charged tracks 
must originate from the vicinity of the $e^+e^-$ 
interaction point. Charged kaon and pion candidates, with
the exception of the slow pion from the decay of the $D^{*+}$, are 
required to have 
ionization losses in the drift chamber within 3.0 and 2.5 
standard deviations, respectively, of those expected for the 
hypothesis under consideration.  The invariant mass of the two photons 
from $\pi^0 \to \gamma\gamma$ must be within 2.0 standard
deviations ($\sigma = 5$ MeV/$c^2$ to 8 MeV/$c^2$, depending on shower 
energies and polar angles) of the nominal $\pi^0$ mass.

The  $K^-\pi^+$ and $K^-\pi^+\pi^0$ combinations are required to have
an invariant mass within 16 MeV/$c^2$ and 25 MeV/$c^2$  ($\sim 2\sigma$) 
of the nominal $D^0$ mass, respectively.
In addition, we select regions of the $D^0 \to K^-\pi^+\pi^0$ Dalitz plot
to take advantage of the known resonant substructure~\cite{E691}, and
we enforce a minimum energy for the $\pi^0$. In the $D^0 \to K^-\pi^+\pi^0$ 
mode we require $|{\bf{p}}_D| > 0.8$  GeV/$c$ in order to further
reject fake $D^0$ background. We then combine $D^0$ candidates 
with $\pi^+$ candidates to form $D^{*+}$
candidates. The slow pion used to form the $D^{*+}$ must have a 
momentum of at least 65 MeV/$c$.  The reconstructed mass difference
$\delta m = M(D^{0}\pi^+)-M(D^{0})$~\cite{massdef} is required to be 
within 2 MeV/$c^2$ of the known $D^{*+} - D^0$ mass difference~\cite{pdbook}.
The $D^{*+}$ candidate is then combined with an additional $\pi^-$
in the event to form a $D_J^0$ candidate.  The $D_J^0$ candidates 
must have a scaled momentum $x_{D_J} = 
p_{D_J}/\left[E_{\rm beam}^2-M^2(D_J)\right]^\frac{1}{2}<0.5$, the 
kinematic limit from $B$ decays.

These $D_J^0$ candidates are then paired with leptons selected as 
described above to form candidates for 
${B^- \to D_J^0 \ell^- \bar{\nu}_{\ell}}$ decays. To suppress background 
from ${\bar{B{^0}}\to D^{*+} \ell^- \bar{\nu}_{\ell}}$, 
we select $D_J^0~\ell^-$ candidates that are consistent with 
${B^- \to D_J^0 \ell^- \bar{\nu}_{\ell}}$ 
decays, and reject $D_J^0~\ell^-$ candidates that are consistent with 
${\bar{B{^0}}\to D^{*+} \ell^- \bar{\nu}_{\ell}}$. Thus, 
we require  $D_J^0~\ell^-$ candidates to have $
|\cos\theta_{B - D_J\ell}| \leq 1$ and $\cos\theta_{B - D^*\ell} < -1$, 
where
\begin{equation}
\cos\theta_{B - D_J\ell}=\frac{|{\bf{p}}_{D_J\ell}|^2 + |{\bf{p}}_B|^2 
-  |{\bf{p}}_{\nu}|^2}{2|{\bf{p}}_B| |{\bf{p}}_{D_J\ell}|}
\end{equation}
and
\begin{equation}
\cos\theta_{B - D^*\ell}=\frac{|{\bf{p}}_{D^*\ell}|^2 + |{\bf{p}}_B|^2
-  |{\bf{p}}_{\nu}|^2}{2|{\bf{p}}_B| |{\bf{p}}_{D^*\ell}|}.
\end{equation}
Here, $\theta_{B - D_J\ell}$ ($\theta_{B - D^*\ell}$) is the angle 
between ${\bf{p}}_B$ and ${\bf{p}}_{D_J\ell}$ (${\bf{p}}_{D^*\ell}$),
where $|{\bf{p}}_B|$ is the known magnitude of the $B$ momentum, and 
${\bf{p}}_{D_J\ell}$ (${\bf{p}}_{D^*\ell}$)
is the momentum of the $D_J^0~\ell^-$ ($D^{*+}~\ell^-$) candidate.
The magnitude of the neutrino momentum $|{\bf{p}}_{\nu}|$ is inferred 
from energy conservation, using the beam energy for the $B$ meson energy
$E_B$.  To reject uncorrelated background (background from events in which
the $D_J^0$ comes from the $\bar{B}$ and the lepton from the $B$)
we require the $D_J^0$ 
and the lepton to be in opposite hemispheres: $\cos\theta_{D_J\ell} < 0$, 
where $\theta_{D_J\ell}$ is the angle between the $D_J^0$ and the lepton.
The remaining uncorrelated background is negligible.  

The ${B^- \to D_J^0 \ell^- \bar{\nu}_{\ell}}$ signal is identified using the 
mass difference $\delta M_J = M(D^{*+}\pi^-) - M(D^{*+})$. To avoid 
multiple $D_J^0~\ell^-$ combinations per event,
we select the best combination in the event using 
M($\pi^0$), M($D^0$), $\delta m$, and 
$M^2(\nu_{\ell})\simeq M_B^2 + M^2(D_J\ell) - 2 E_B E(D_J\ell)$.
In the computation of $M^2(\nu_{\ell})$, the $B$ meson 
momentum, ${\bf{p}}_B$, is taken to be zero, and $E(D_J\ell)$ is the 
energy of the $D_J^0~\ell^-$ candidate.

\begin{figure}[htb]
\centerline{\psfig{figure=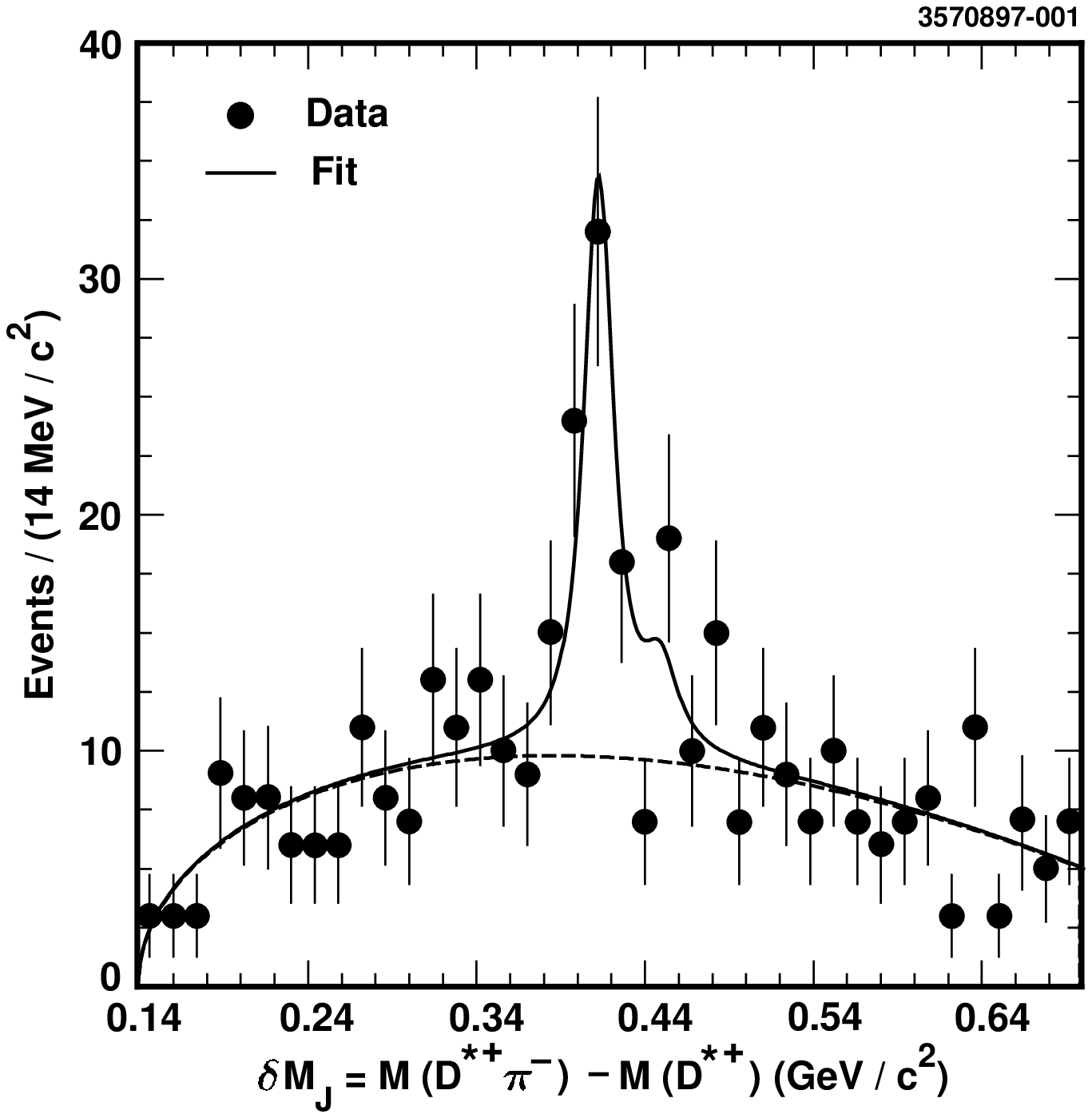,height=7.1in}}
\caption{\baselineskip=0.23in The $\delta M_J$ distribution from 
the $\Upsilon(4S)$ resonance data for ${B^- \to D_1^0 \ell^- \bar{\nu}_{\ell}}$ 
and ${B^- \to D_2^{*0} \ell^- \bar{\nu}_{\ell}}$ ($\ell = e$ and $\mu$)
candidates obtained by combining both the $D^0 \to K^-\pi^+$ and 
$D^0 \to K^-\pi^+\pi^0$ modes. The dashed curve describes the background 
function, whereas the solid line is the sum of the background
and signal functions.}
\label{fig:mass}
\end{figure}

The $\delta M_J$ distribution obtained by combining
the two decay modes of the $D^0$ meson is shown in 
Fig.~\ref{fig:mass}.  An unbinned likelihood fit 
is performed on the $\delta M_J$ distribution. The fitting function
is the sum of a threshold background function~\cite{backfunc} plus 
Breit-Wigner resonance functions with the masses and widths of 
the two narrow $D_J^0$ resonances fixed~\cite{pdbook}. Each Breit-Wigner 
function is convoluted with a Gaussian that describes the detector 
resolution.  The width of the Gaussian is estimated by Monte Carlo 
simulation to be $\sigma$ = 2.8 MeV/$c^2$.  The $D_1^0$ and $D_2^{*0}$ 
yields obtained from the fit are summarized in Table~\ref{table:one}.

To check that the data are consistent with the presence of a signal, we 
fit the $\delta M_J$ distribution with only the 
smooth background function. The difference between the logarithm of the 
likelihood of the fit with the signal plus the background functions 
and the logarithm of the likelihood with only the background 
function is 18.7. Assuming Gaussian statistics, this corresponds to a 
$6.1 \sigma$ statistical significance of the signal over the 
background. If the mass and the width of the $D_1^0$ resonance are 
allowed to float, the fitted mass and width obtained are
$2420 \pm 4$ MeV/$c^2$ and $23 \pm 9$ MeV/$c^2$; which is
in agreement with the PDG averages~\cite{pdbook}.
The $D_1^0$ and $D_2^{*0}$ yields from this fit are $62.5 \pm 16.7$ and 
$10.5 \pm 9.8$, respectively.

The background from non-$B\bar{B}$ events is obtained by 
measuring the signal yields using OFF Resonance data. The 
results are scaled 
by the ratio of the luminosities and the square of the beam energies.
Fake lepton background (the contribution in which a $D^0_J$ is paired
with a hadron misidentified as a lepton)
is estimated by performing the same analysis
using tracks that are not leptons. The fake lepton yields are scaled by the 
appropriate misidentification hadron probabilities and abundances.
The sums of these two types of backgrounds are subtracted from the
ON Resonance Yields as indicated in Table~\ref{table:one}.

\begin{table}[t]
\caption{\label{table:one} Yields and product branching
fractions.  The first error on the product branching fractions is 
statistical,  the second is experimental systematic and the third is 
theoretical.}
\begin{center}
\begin{tabular}{ccc}
                           & $D_1^{0}$             & $D_2^{*0}$         \\
\hline
ON Resonance Yield         & $56.6 \pm 11.9$       &  $10.3 \pm 9.4$    \\
Background Yield           & $ 3.1 \pm  2.8$       &  $ 1.5 \pm 2.8$    \\
\hline
Net Yield                  & $53.5 \pm 12.2$       &  $ 8.8 \pm 9.8$    \\
\hline
${\cal P}(D_J^0)$   & $(0.373 \pm 0.085 \pm 0.052 \pm 0.024)$ \%
                 & $(0.059 \pm 0.066 \pm 0.010 \pm 0.004)$ \%       \\
\end{tabular}
\end{center}
\end{table}

Semileptonic $\bar{B}$ decays to more highly excited charmed mesons which
then decay to $D_J^0$ mesons are predicted to be small \cite{isgw2}.
The smooth background function includes both combinatoric background and 
background from broad and non-resonant $D^{*+} \pi^- X$ states.

The product branching fractions ${\cal P}(D_1^0) = 
{\cal B}(B^- \to D_1^0 \ell^- \bar{\nu}_{\ell}){\cal B}(D_1^0 \to D^{*+}\pi^-)$ 
and ${\cal P}(D_2^{*0}) = {\cal B}(B^- \to D_2^{*0} \ell^- \bar{\nu}_{\ell})
{\cal B}(D_2^{*0} \to D^{*+}\pi^-)$ are obtained by dividing the yields 
by the total numbers of $B^-$ events in our data sample and the sum of 
the products of the efficiencies times the $D^{*+}$ and $D^0$ branching 
fractions for the modes used.  The reconstruction efficiencies ($\varepsilon_{D_J}$) 
for $B^- \to D_J^0 \ell^- \bar{\nu}_{\ell}$ ($\ell = e$ or $\mu$)
are $\varepsilon_{D_1}^{K\pi} = (4.37 \pm 0.93)\%$,
$\varepsilon_{D_1}^{K\pi\pi^0} = (1.09 \pm 0.02)\%$, $\varepsilon_{D_2^*}^{K\pi} 
= (4.61 \pm 0.97)\%$, and $\varepsilon_{D_2^*}^{K\pi\pi^0} = (1.10 \pm 0.02)\%$.
Our event selection 
efficiencies were obtained using Monte Carlo data generated according 
to the ISGW2 model~\cite{isgw2}.  We assume that the branching
fractions of $\Upsilon(4S)$ to charged and neutral $B\bar{B}$ pairs 
are each 50\%.  The values of the $D^{*+}$ and $D^0$ branching 
fractions are taken from Ref.~\cite{pdbook}.  The contributions of 
the systematic uncertainties are listed in Table~\ref{table:two}.
Details on the systematic uncertainties estimation
can be found elsewhere~\cite{thesis}.  The theoretical uncertainties 
associated with the model dependence of the efficiency is obtained by 
varying the parameters and the form factors used in the ISGW2 model.  
We choose to quote the product of branching fractions because the 
branching fractions for $D^0_J \to D^{*+}\pi^-$ have not been measured. 
We find:
\begin{eqnarray}
\label{eq:br}
{\cal P}(D_1^0)    
&=& (0.373 \pm 0.085 \pm 0.052 \pm 0.024)~\% \\
{\cal P}(D_2^{*0}) 
&=& (0.059 \pm 0.066 \pm 0.010 \pm 0.004)~\% \nonumber \\
&<& 0.16~\%~(90\%~{\rm C.L.}),
\end{eqnarray}
where the errors are statistical, systematic and theoretical, 
respectively.  For the quoted upper limit, we add the 
experimental systematic and the theoretical uncertainties in quadrature, 
and add the result to the upper limit computed with the statistical 
error only.

The uncertainties on the widths of the $D_J^0$ resonances turn out 
to be our biggest systematic uncertainty. Fortunately, the dependence 
of ${\cal P}(D_1^0)$ on the width of the $D_1^0$ can be 
parameterized:
\begin{eqnarray}
{\cal P}(D_1^0) &=& ({\cal P}(\Delta\Gamma) \pm 0.085 \pm 0.037 \pm 0.024)~\%,
\end{eqnarray}
where ${\cal P}(\Delta\Gamma) = (0.373 + 9.25\times10^{-2} \, \Delta\Gamma)\%$, 
with $\Delta\Gamma = \Gamma - \Gamma_0$ ($\Gamma_0 = 
18.9$ MeV/$c^2$~\cite{pdbook}). The value of the slope
$d{\cal P}/d\Gamma=9.25\times10^{-2}$ MeV$^{-1}c^2$ is extracted from a 
linear fit of ${\cal P}(D_1^0)$ versus $\Delta\Gamma$.

\begin{table}[t]
\caption{\label{table:two} Experimental systematic 
errors on the product branching fractions.  Tracking uncertainties are 
for all charged particles other than the slow $\pi$.}
\begin{center}
\begin{tabular}{ccc}
Source of                           & ${\cal P}(D_1^0)$ & ${\cal P}(D_2^{*0})$ \\
Systematic Error                    &                   &                      \\
\hline
$M_{D_J}$                            &   1.0\%           &  1.1\%  \\
$\Gamma_{D_J}$                       &  10.0\%           & 14.0\%   \\
Background Function                  &   4.0\%           &  5.0\%  \\
Uncorrelated Background              &   0.5\%           &  0.4\%   \\
Lepton Fake                          &   1.0\%           &  1.0\%   \\
Lepton ID                            &   1.3\%           &  1.3\%   \\
MC Statistics                        &   1.5\%           & 1.5\%  \\
${\cal B}(D^{*+} \to D^0 \pi^+)$     &   2.0\%           & 2.0\%  \\
${\cal B}(D^0 \to K^- \pi^+(\pi^0))$ &   3.5\%           & 3.5\%  \\
Slow $\pi$ Efficiency                &   5.0\%           & 5.0\%  \\
Tracking Efficiency                  &   4.0\%           & 4.0\%   \\
$\pi^0$ Reconstruction               &   2.4\%           & 2.4\%  \\
Dalitz Weight                        &   1.9\%           & 1.9\%  \\
Multiple Counting                    &   1.4\%           & 1.4\%  \\ 
Particle Identification              &   1.0\%           & 1.0\%  \\
Luminosity                           &   2.0\%           & 2.0\%  \\
\hline
Total                                &  14.0\%           & 17.3\% \\
\end{tabular}
\end{center}
\end{table}

In order to estimate the contribution of these decays to the total
semileptonic $B$ meson branching fraction, we need
to make some assumptions about the branching fractions of the
$D_J^0$ mesons.
Isospin conservation
and CLEO measurements~\cite{cleodj}
of the decays of the $D_J^0$ mesons suggest
that ${\cal B}(D_1^0 \to D^{*+}\pi^-) = 67\% $ and ${\cal B}(D_2^{*0}
\to D^{*+}\pi^-) = 20\% $.  Using these estimates we find
\begin{eqnarray}
{\cal B}(B^- \to D_1^0 \ell^- \bar{\nu}_{\ell})    
&=& (0.56 \pm 0.13 \pm 0.08 \pm 0.04)~\% \\
{\cal B}(B^- \to D_2^{*0} \ell^- \bar{\nu}_{\ell}) 
&<& 0.8~\%~~ (90\%~{\rm C.L.}),
\end{eqnarray}
where no attempt has been made to estimate the systematic uncertainties 
due to the $D_J^0 \to D^{*+}\pi^-$ branching fractions.

A clear picture of the exclusive modes which make up 
the 30 -- 40\% of the $B$ semileptonic decays that are not
$D\ell\nu$ and $D^*\ell\nu$ has not yet emerged. However, it appears 
that no more than half of the excess can be due to exclusive 
semileptonic decays to $D_1^0(2420)$ and $D_2^{*0}(2460)$.

In summary, we have studied exclusive semileptonic
decays of the $B$ mesons to P-wave charm mesons.
We measured a branching fraction for 
${\cal B}(B^- \to D_1^0 \ell^- \bar{\nu}_{\ell})
{\cal B}(D_1^0 \to D^{*+}\pi^-)$ and
an upper limit for ${\cal B}(B^- \to D_2^{*0} \ell^- \bar{\nu}_{\ell})
{\cal B}(D_2^{*0} \to D^{*+}\pi^-)$.
These results indicate that a substantial fraction 
(${\raisebox{-.65ex}{\rlap{$\sim$}} \raisebox{.45ex}{$>$}} 20\%$) of the
inclusive
$B$ semileptonic rate is from modes other than $D\ell\nu$, $D^*\ell\nu$,
$D_1 \ell \nu$ and $D_2^* \ell \nu$.

We gratefully acknowledge the effort of the CESR staff in providing us with
excellent luminosity and running conditions.
This work was supported by
the National Science Foundation,
the U.S. Department of Energy,
the Heisenberg Foundation,
the Alexander von Humboldt Stiftung,
the Natural Sciences and Engineering Research Council of Canada, 
le Fonds Qu\'eb\'ecois pour la Formation de Chercheurs et 
l'Aide \`a la Recherche, and the A.P. Sloan Foundation.

\nopagebreak


\newpage

\begin{thebibliography}{99}
\bibitem{chgconj} {Here and throughout the paper, charged conjugate
modes are implied.}
\bibitem{massdef} {The invariant mass of the particles included 
in a candidate for the state $X$ is denoted by $M(X)$.}
\bibitem{backfunc} {The background function used is $f(x) = a_1 \sqrt{x} 
\times (1 + a_2 x + a_3 x^2)$, where the $a_i$ are free 
parameters and $x = \delta M_J - m_{\pi^-}$.}
\end{thebibliography}
\end{document}